\documentclass[prb,showpacs,twocolumn,aps,a4paper]{revtex4}
\usepackage{amsmath}
\usepackage{graphicx}
\usepackage{latexsym}
\usepackage{amsfonts}
\usepackage{amssymb}
\DeclareGraphicsExtensions{.pdf,.gif,.jpg}

\newcommand{\be}{\begin{equation}}
\newcommand{\ee}{\end{equation}}
\newcommand{\beq}{\begin{eqnarray}}
\newcommand{\eeq}{\end{eqnarray}}

\begin{document}

\def\bbe{\mbox{\boldmath $e$}}
\def\bbf{\mbox{\boldmath $f$}}
\def\bg{\mbox{\boldmath $g$}}
\def\bh{\mbox{\boldmath $h$}}
\def\bj{\mbox{\boldmath $j$}}
\def\bq{\mbox{\boldmath $q$}}
\def\bp{\mbox{\boldmath $p$}}
\def\br{\mbox{\boldmath $r$}}

\def\bone{\mbox{\boldmath $1$}}

\def\dr{{\rm d}}

\def\tb{\bar{t}}
\def\zb{\bar{z}}

\def\tgb{\bar{\tau}}

\def\bC{\mbox{\boldmath $C$}}
\def\bG{\mbox{\boldmath $G$}}
\def\bH{\mbox{\boldmath $H$}}
\def\bK{\mbox{\boldmath $K$}}
\def\bM{\mbox{\boldmath $M$}}
\def\bN{\mbox{\boldmath $N$}}
\def\bO{\mbox{\boldmath $O$}}
\def\bQ{\mbox{\boldmath $Q$}}
\def\bR{\mbox{\boldmath $R$}}
\def\bS{\mbox{\boldmath $S$}}
\def\bT{\mbox{\boldmath $T$}}
\def\bU{\mbox{\boldmath $U$}}
\def\bV{\mbox{\boldmath $V$}}
\def\bZ{\mbox{\boldmath $Z$}}

\def\bcalS{\mbox{\boldmath $\mathcal{S}$}}
\def\bcalG{\mbox{\boldmath $\mathcal{G}$}}
\def\bcalE{\mbox{\boldmath $\mathcal{E}$}}

\def\bgG{\mbox{\boldmath $\Gamma$}}
\def\bgL{\mbox{\boldmath $\Lambda$}}
\def\bgS{\mbox{\boldmath $\Sigma$}}

\def\bgr{\mbox{\boldmath $\rho$}}

\def\a{\alpha}
\def\b{\beta}
\def\g{\gamma}
\def\G{\Gamma}
\def\d{\delta}
\def\D{\Delta}
\def\e{\epsilon}
\def\ve{\varepsilon}
\def\z{\zeta}
\def\h{\eta}
\def\th{\theta}
\def\k{\kappa}
\def\l{\lambda}
\def\L{\Lambda}
\def\m{\mu}
\def\n{\nu}
\def\x{\xi}
\def\X{\Xi}
\def\p{\pi}
\def\P{\Pi}
\def\r{\rho}
\def\s{\sigma}
\def\S{\Sigma}
\def\t{\tau}
\def\f{\phi}
\def\vf{\varphi}
\def\F{\Phi}
\def\c{\chi}
\def\w{\omega}
\def\W{\Omega}
\def\Q{\Psi}
\def\q{\psi}

\def\ua{\uparrow}
\def\da{\downarrow}
\def\de{\partial}
\def\inf{\infty}
\def\ra{\rightarrow}
\def\lra{\leftrightarrow}
\def\bra{\langle}
\def\ket{\rangle}
\def\grad{\mbox{\boldmath $\nabla$}}
\def\Tr{{\rm Tr}}
\def\Re{{\rm Re}}
\def\Im{{\rm Im}}

\title{Quantum Rings in Magnetic Fields and Spin Current Generation }

\author{Michele Cini}

 \affiliation{ Dipartimento di Fisica, Universit\`{a}
di Roma Tor Vergata, Via della Ricerca Scientifica 1, I-00133 Rome,
Italy, and Istituto Nazionale di Fisica Nucleare - Laboratori
Nazionali di Frascati, Via E. Fermi 40, 00044 Frascati, Italy.}

\author{Stefano Bellucci}
 \affiliation{  Istituto Nazionale di Fisica Nucleare - Laboratori
Nazionali di Frascati, Via E. Fermi 40, 00044 Frascati, Italy.}
\date{\today}

\begin{abstract}
We propose three different mechanisms for pumping spin-polarized currents in a ballistic circuit using a time-dependent magnetic field acting on an
 asymmetrically connected quantum ring at half filling.   The first mechanism works thanks to a  rotating magnetic field and produces an alternating current
 with a partial spin polarization.  The second  mechanism works by  rotating the ring in a constant  field; like the former case, it  produces an
 alternating charge current but the spin current is d.c.; both methods  do not require a spin-orbit interaction to achieve the polarized current, but the rotating ring could be used to
 measure the spin-orbit interaction in the ring using characteristic oscillations. On the other hand, the last mechanism that we propose  depends  on the spin-orbit interaction in an essential way, and requires  a  time-dependent magnetic field in the plane of the ring. This arrangement  can  be designed to pump a purely spin current.  The absence of a charge current  is demonstrated   analytically. Moreover, a simple formula for the current is derived and compared to the numerical results.
\end{abstract}


\maketitle

\section{Introduction and background}

Spintronics is an ambitious project which is still largely
hypothetical but is making fast progress  in connection with
potential applications in memory devices, optoelectronics,  and,
among others, quantum information processing \cite{devices1,
devices2, devices3, devices4}. The most  basic question is  how to
generate the spin currents and inject them efficiently  into the
envisaged spintronic  circuits. Johnson and Silsbee\cite{js} in a pioneering work excited a spin current from ferromagnetic electrods into Al stripes.  Now there are several other proposals. One is based on adiabatic modulations of  magnetic stripes\cite{benjamin1};  another adiabatic proposal is
based on the competition between normal and Andreev reflections\cite{benjamin2}. Early   methods to (partially) polarize currents, like the Datta-Das
 transistor\cite{dattadas}, use a two-dimensional electron gas and an
 effective magnetic field due to the spin-orbit interaction.  Indeed it can be seen e.g. in the review by Shen \cite{shen} that  most approaches  for producing  spin polarized currents are based on the spin-orbit interaction. The effectiveness of the process and the degree of polarization  depends on the Rashba interaction, which  has been measured at  low temperatures in InAs,  one of the promising
 materials\cite{inas}. Quite recently,  Sadreev and Sherman explored the possibility of controlling
 the spin-flip conductivity in a wire; they found that this is feasible  by crossed electric and magnetic
 fields exploiting the Rashba spin-orbit effect\cite{sadreev}. It was
 shown that by modulating the transversal electric field one can produce
 some spin polarization in the current through the wire.

The spin current can be measured \cite{shen}. Remarkably, not only  most routes  for spin current  generation but also detection methods are also based on the spin-orbit interaction which can convert spin currents to charge currents\cite{cui}.

Here we focus on the possible use of rings, which for fundamental topological reasons offer a complementary and also in part  an alternative to the mechanism driven by the  spin-orbit interaction  to polarize the current  and send it across a circuit.   We wish to present several novel ways to produce and pump highly polarized spin currents  based on the use of normal metal rings.
Incidentally, a recent paper\cite{last} already deals with a  one-dimensional quantum ring
containing few electrons,  endowed with a Rashba interaction,
linear in the electron momentum. The result is that the  Rashba effects, which is a part of the spin-orbit coupling, generates a pure spin
current\cite{last};  however  the spin current is confined in the ring
 since there are no wires in the model.

We are interested in connected rings, instead.
Besides producing the spin polarized current, one must solve the
problem of efficiently injecting this current into the spintronic
circuit. Classically the spin currents are of course unknown and
at any rate there is no way to excite a current in a circuit
without  using a bias across its ends. In mesoscopic or ballistic
conductors, however, quantum mechanics produces  several kinds of
pumping phenomena that have been highlighted  by several
authors.\cite{pump1, pump2, pump3, pump4, pump5}.

 A
preliminary question naturally arises: {\em is it possible in
principle to  use rings to achieve pumping?} The answer is {\em
yes} for nontrivial reasons rooted into the nonlinear quantum
nature of the spin magnetic moment of nano-rings. This statement
needs an immediate explanation and the reasoning runs as follows.

A  planar connected ring has a magnetic moment which
develops when a current is excited by a bias across the circuit, provided that the connection to the circuit breaks parity in the plane.
Classically the magnetic moment is linear in the current and in
the bias\cite{jackson}.  Consider now the the ring in the absence
of a bias. The effect of a magnetic flux in the ring is to excite
a current in its sides. Provided the flux also pumps current in
the external  circuit, taking a net amount of charge from a wire
to the other, then  one has an example of one-parameter pumping.
However the Brouwer theorem\cite{brower} forbids one-parameter
pumping  in a linear system. At this stage of the reasoning the
answer to the above question is definitely {\em no}. This
prediction is reversed by the finding\cite{ciniperfettostefanucci}
that the magnetic moments excited by currents in nanoscopic
circuits containing loops are dominated by quantum effects and
depend nonlinearly on the exciting bias, quite at variance from
classical expectations of a linear behavior.

Now we expect that ring can be used for pumping under suitable
conditions and this has been verified\cite{ciniperfetto} by
studying the real-time quantum evolution of tight-binding models
in different geometries. It yields a nonadiabatic pumping. An
arbitrary amount of charge can thereby be transferred from one
side to the other by suitable flux protocols. Furthermore, in
order to take electron-electron interactions into account, a
quantum ring laterally connected to open one-dimensional leads was
described within the Luttinger liquid model. \cite{ll5} The
interactions do not hamper the pumping effect. By a different
arrangement one can model a memory device, in which both
operations of writing and erasing can be performed efficiently and
reversibly.  As a direct consequence of the above mentioned
nonlinearity, one can achieve, by employing suitable flux
protocols, single-parameter nonadiabatic pumping.

 The above findings  hold for spinless models but here we wish to show
 that they open new directions in the search for practical methods to
 manipulate spins and spin currents. Thus, the  present paper aims at
 extending the theory to include the spin- field and spin-orbit
 interactions and looking for new means for producing and pumping spin
 currents. We consider  laterally connected normal metal rings with any
 even number of sides; the restriction to even numbers has a fundamental
 mathematical reason that will be apparent in Sect. VI below. Ring and
 leads are taken to be at half filling, (unlike the ring of
Ref\cite{last}), with no external bias.

 The advantage of the present approach is that it  starts from different
 topology, or genus; we can achieve high and even total spin polarization
 by three different
 methods that can work even at room temperature.  The results depend
 in an essential way on the spin
magnetic interaction, whereas one of them can also do without the
spin-orbit coupling. The effects of the spin-orbit interactions on
the above thought experiments are also explored and play a crucial
role in the last arrangement. The conditions for a maximally spin
polarized current are presented below in all cases. The spin
polarizers that we propose are based on the use of a normal (i.e.
neither superconducting nor magnetic) metal with half-filled
bands.

The plan of the paper is as follows.  Sect. II is devoted to the
geometry and formalism, Sect. III explains our research strategy.
Sect. IV describes the spin current generated by a rotating
magnetic field. In Sect. V it is the ring that rotates, keeping
the spin quantization axis fixed. Sect. VI describes the pure spin
current obtained by a variable field in the plane of the ring; the
proof that the current is a pure spin current is given in
Appendix. Sect. VII presents our conclusions.

\begin{figure}[h]
\includegraphics*[width=.30\textwidth]{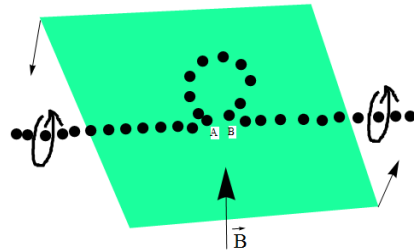}
\caption{We consider laterally connected rings, that are maximally asymmetric. In Sect.IV we consider the ring fixed in a rotating magnetic field; in Sect. V, we show that if the ring rotates around the AB bond in a fixed magnetic field,  as depicted here, the time-dependent  flux produces a spin-polarized current. In Sect.VI instead we study the effect of a tangential magnetic field. }
\end{figure}

\begin{figure}[h]
\includegraphics*[width=.50\textwidth]{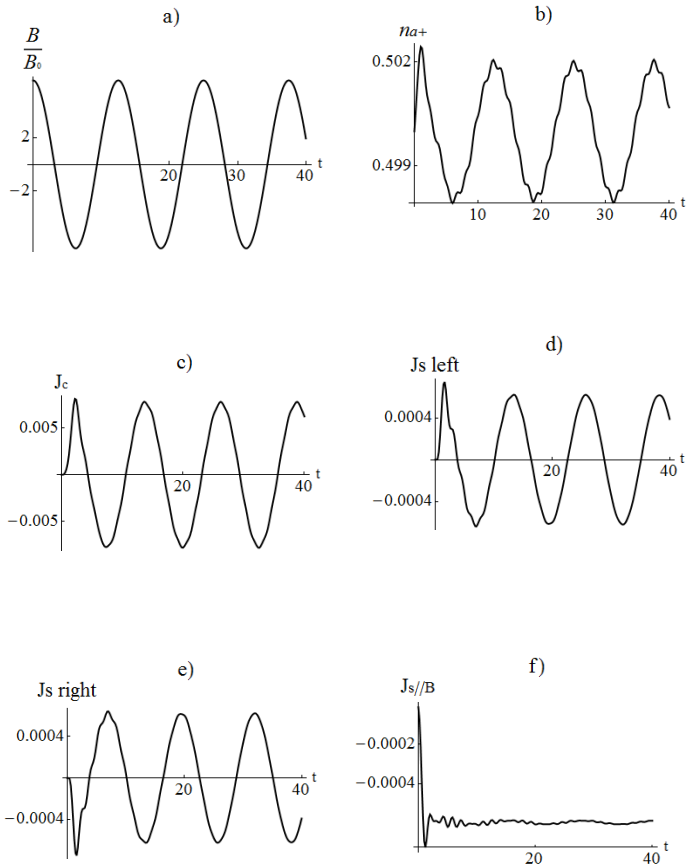}
\caption{Here we consider a $N_{ring}=250$ atom ring without spin-orbit interaction.  Time $t$ is measured in $\frac{\hbar}{t_{h}}$ units. In Figures 2a) -2e)the ring is fixed  in an external rotating magnetic field, with $\omega=\frac{2\pi t_{h}}{12.5\hbar}$.  Figure 2a) shows normal component  $B_{\lfloor}(t))$ of the magnetic field in units of   $B_{0}=\frac{hc}{Se}$, where $S=\frac{N_{ring}a^{2}}{4 \tan(\frac{\pi}{N_{ring}})}$ is the surface of the ring. So,  $B_{\lfloor}$ produces a fluxon in the ring.  Fig. 2b): the spin up  occupation of site A, where the ring is joined to the left lead. Figure 2c):  charge current on the first bond of the left lead. Figure 2d):  spin current on the first bond of the left lead,  polarized normal to the ring. Figure 2d):  spin current on the first bond of the left lead,  polarized normal to the ring. Figure 2e):  spin current on the first bond of the right lead,  polarized normal to the ring; the spin current has a negative parity. In Figure 2f) the ring is rotating with angular frequency  $\omega$ in a constant  external magnetic field. and we show the spin current on the first bond of the left lead,  polarized  parallel to the magnetic field $B$. }
\end{figure}

\begin{figure}[]
\includegraphics*[width=.50\textwidth]{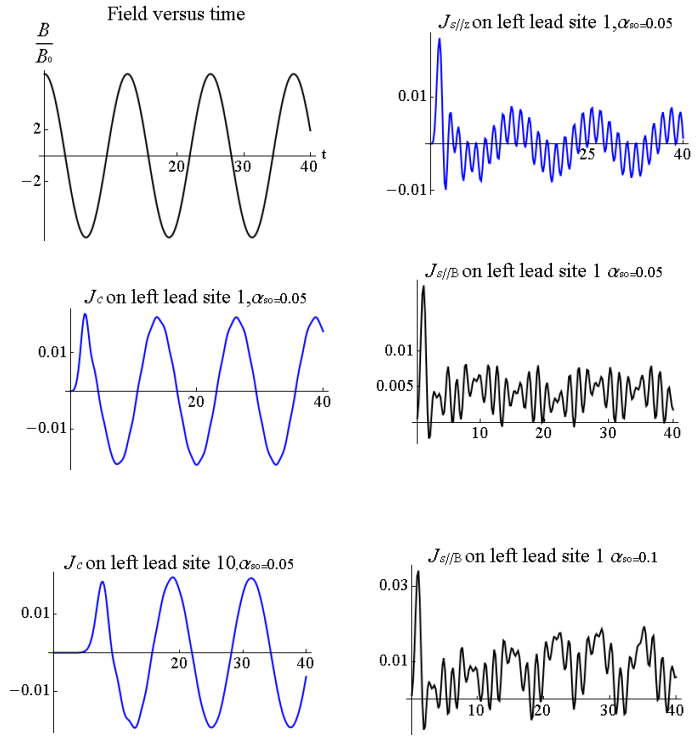}
\caption{Here we consider a 100 atom ring  rotating in an external magnetic field with $\omega=\frac{2\pi t_{h}}{12.5\hbar}$. The  spin-orbit interaction producing a phase shift  ($\alpha_{SO} =0.05$). Top left:  the field as in Fig.2  versus  time (in $\frac{\hbar}{t_{h}}$ units. Middle left:  charge current on the first bond of the left lead. Bottom left:  charge current on the tenth bond of the left lead. Top right:
 spin current on the first bond of the left lead,  polarized normal to the ring. A   modulation at frequency higher then $\omega$  is due to the spin-orbit interaction.  Middle right:spin current on the first bond of the left lead,  polarized  parallel to the magnetic field $B$.Compared to Fig. 2e) the spin current is much larger, with wide oscillations. The difference is a size effect combined with an  effect of the spin-orbit coupling. Bottom right: the same except that  $\alpha_{SO} =0.1$ }
\end{figure}

\section{Laterally Connected Ring in Magnetic Field}

The three thought experiments that we wish to perform use  a ring shaped as a regular polygon with $N_{ring}$ sides, laterally connected to wires, plunged in an external magnetic field. The geometry of the laterally connected ring is sketched in Fig.1 for $N_{ring}=10$ and in Fig. 4a) for $N_{ring}=8.$   Since the effects that we introduce below are topological the precise value of a and the geometrical details are not crucial; taking for simplicity a regular polygon, the surface area is $$S=\frac{N_{ring}a^2}{4 \tan(\pi /N_{ring})},$$
where the side length $a$ is of the order of one Angstrom.  In the
ring the electrons feel the field and a spin-orbit interaction.
There is no bias applied to the circuit; the time-dependent
external magnetic field $\overrightarrow{B}$ acting on the ring  has a component
$B_{\shortparallel}$ in the plane of the ring and a normal component
$B_{\lfloor}$ which produces a flux $\alpha=B_{\lfloor} S.$ For the sake of definiteness we
assume  that each bond in the ring is modified by the field
according to the symmetric prescription
$t_{ring}\rightarrow t_{ring}\exp[{i\alpha(t)/N_{ring}}],$ although the symmetry is not important here.

  We outline the above situation by the tight-binding model  Hamiltonian
\be H=H_{D}+H_{B},\ee where $H_{D}$ describes  the device  and
$H_{B}$ the magnetic term.  \be
H_{D}=H_{ring}+ H_{wires}+H_{ring-wires}.\ee The $N_{ring}$-sided
ring, with even $N_{ring}$, lies in the $x-y$ plane and   is represented by \be
 H_{ring}=t_{ring} e^{ i \frac{\alpha(t)}{N_{ring}}}\sum_{\sigma=\pm 1}\exp{[{i\sigma\alpha_{SO}}]}c^{\dagger}_{i+1,\sigma}c_{i,\sigma}+h.c.\ee where as usual $c_{i,\sigma}$ annihilates an electron  at site $i$ of spin projection $\sigma\frac{\hbar}{2}$, the quantization axis being the z axis.
 We also use the common pictorial notations $c_{i,\uparrow}=c_{i,+}$ and $c_{i,\downarrow}=c_{i,-}.$
  The  further magnetic interaction due to $B_{\shortparallel}$ acts exclusively on spin and is taken to be:
\be H_{B}=V(t) \sum_{i \in
ring}(c^{\dagger}_{i,\uparrow}c_{i,\downarrow}+
c^{\dagger}_{i,\downarrow}c_{i,\uparrow}   )\ee where $V=\mu_B B$
with the Bohr magneton $\mu_B=5.79375 *10^{-5}\frac{eV}{Tesla}.$

The form of the spin-orbit interaction is borrowed from A. A.
Zvyagin\cite{Zvyagin}, where it is represented as a phase  shift
$\alpha_{SO}$ for up-spin  and $-\alpha_{SO}$ for down-spin
electrons. This implies that opposite spin electrons feel opposite
effective fluxes in the ring. In this way the $z$ direction
becomes a privileged one in spin space.

 In order to model the Left and Right  wires, that extend, say,  along the x axis, we introduce the corresponding creation and annihilation operators and write:
 \begin{equation} H_{wires}=\sum_{\eta=Left,Right}H_{\eta}, \end{equation} where \begin{equation}H_{\eta}=t_{h}\sum_{n,\sigma} c^{\dagger^{(\eta)}}_{n,\sigma}c^{(\eta)}_{n+1,\sigma}+h.c.,
  \end{equation}
  The ring-wires contacts consist of hopping integrals  $t_{lr}$ that connect two nearest-neighbor sites of the ring denoted with  A and B with the first sites of lead L and R, respectively.

 In the Hamiltonian they are represented by a term $$H_{leads-ring}=t_{lr}(\sum_{\sigma}c^{\dagger Left}_{1,\sigma} c_{A,\sigma}+\sum_{\sigma}c^{\dagger Right}_{1,\sigma} c_{B,\sigma})+h.c.$$
Below we
assume for the sake of definiteness  that $t_h=t_{ring}=t_{lr}=1$ eV. \\

  The equilibrium  occupation of the system (which prevails for $t<0$) is
  determined by the spin-independent chemical potential $\mu_c$.
  In this paper we consider $\mu_c=0$ (i.e. we assume half filling).

In order to describe its evolution of the system starting from its
ground state, we begin our consideration from the retarded Green's
function matrix elements in a spin-orbital basis: \be
g^{r}_{i,j}=\langle i|U_I (t,0)|j\rangle \ee where $U_I (t,0)$ is
the evolution operator in the interaction representation. Taking
the spin quantization axis along z,  the number
current\cite{cini80} is \be J_{n,\sigma} (t)=-2 \frac{t_h}{h} Im (
G^{<}_{n,\sigma,n-1,\sigma})\ee\label{z} where \be
G^{<}_{i,j}(t)=\sum_\mu n^{0}_\mu g^{r}_{i,\mu}(t,0)g^{r
*}_{j,\mu}(t,0).
\end{equation}
where $\mu$ denotes the ground state spin-orbitals for $B=0$ and $n^{0}_\mu$ is the Fermi function.
The tight-binding model is a simple but often useful approximation; one advantage is that it allows exact results, as those concerning the pure spin current derived in the Appendix.
\subsection{Units used in the numerical calculations}
 We shall simulate several  experimental settings. In all cases, our codes calculate number currents taking the hopping integral  $t_h =1.$  For instance if  this is interpreted to mean
that $t_h =1 $eV, which corresponds to the frequency $2.42
*10^{14}$ $s^{-1},$ a current $J=1$ from the code means $2.42
*10^{14}$ electrons per second, which corresponds to a charge
current of  $3.87*10^{-5}$ Ampere.

\section{The physical problem: spin-dependent pumping}

We may start our reasoning from  the study of the spinless
case\cite{ciniperfetto} where adding a flux quantum to the flux piercing
the ring is a gauge transformation. While the  Hamiltonian is left
invariant, the transient electric field pumps  a finite amount of
charge towards the left or the right wire according to the sign of
the flux. The spin magnetic moment spoils the invariance of the
Hamiltonian producing an  energy shift between up and down spins
(quantized normal to the ring plane), and any pumping mechanism
will now be spin-polarized. Since the field needed to produce a
fluxon is inversely proportional to the surface, the spinless
model\cite{ciniperfetto} continues to work approximately for large enough
rings while it breaks down for small ones.  Below we consider
three possibilities to overcome this; the first two are  based on
the idea that in a cyclic process up and down spins are again put
on equal footing, while the last works with a field in the ring
plane, which again treats up and down spins (quantized normal to
the ring plane) on equal footing. Finally,in the present analysis we take the bands to be half filled, for the sake of having all the possible symmetry, and the ring is taken with an even number of sides, since in this way the whole system is bipartite. This in particularly important in Sect. 6.

\section{Rotating magnetic field}

In the first arrangement that we consider in this work we leave
the ring fixed in the x-y plane while the field  rotates; in Cartesian coordinates,
  $\overrightarrow{B}(t)=(0,B_{\shortparallel}(t),B_{\lfloor}(t))$, with $B_{\shortparallel}=B \cos(\omega t)$ and  $B_{\lfloor}=B
\sin(\omega t).$ In Figures 2a) to 2d) we report the results of  a
simulation with a 250 atom ring and $\omega=\frac{2\pi
t_{h}}{12.5\hbar}$ in which $\alpha_{SO}=0,$ i.e. no spin-orbit interaction is included.  The
magnetic flux has a sine dependence on time shown in Figure 2a).
 In the numerical code the wires are 50 atoms
long, which is enough to simulate infinite wires for the time
interval up to $40 \frac{\hbar}{t_h}$ that we are considering, since longer wires give the same results.

Fig. 2b) shows the time dependence of the spin up occupation of site A,
where the ring is joined to the left lead. The charge is seen to
oscillate without showing a trend to deviate steadily  from half
filling. In Figure 2c) we see a similar oscillatory trend of the
charge current on the first bond of the left lead.

   Figures 2d)and 2e)
show the spin current on the first sites of the left ant right lead, respectively;
the spin quantization axis is  in the $z$ direction orthogonal  the ring. As above,
$J=1$ from the code means $2.42
*10^{14}$ spins per second, if $t_h =1 $eV, and the spin current is proportional to $t_h$.

 We conclude that a  rotating magnetic field around a fixed ring
would pump an alternating spin current in the wires. The driving
force can be understood in terms of the contents of the previous
Section. At this stage, the results are already  striking:
an oscillating charge current is to be expected even classically in the ring, but the pumping is a purely quantum mechanical phenomenon.
Once the sense of rotation of the magnetic field is chosen,
clockwise and anticlockwise  are made physically different and up
and down spins have different energies in the adiabatic limit; it
is clear that this situation produces the alternating
charge current.  Moreover, one  can see that changing the sense
of rotations all currents must change sign, as the numerical
solution shows. In addition, this reasoning suggests that the B
dependence must be essentially linear and that the spin-orbit
interaction does not play an  essential role. However, pumping has already been described in a spin-less model\cite{ciniperfetto}. The most important novelty is the spin current, which is obtained without the help of the spin-orbit  interaction, which has always been used to polarize currents in previous works, as mentioned above. The new ingredient is the topology, which {\em per se} allows to produce spin currents. An even more striking
possibility comes next.

 \section{Rotating ring}

 The second  arrangement that we consider in this work is shown in Figure 1, where the ring  rotates in a constant  field. In the  reference of the ring the field   still rotates according to $B_{\shortparallel}=B \cos(\omega t)$ and  $B_{\lfloor}=B \sin(\omega t),$ as in the previous case. The difference is that now the spin quantization axis is kept parallel to the magnetic field $B$. The  current $J_{m}(\overrightarrow{n})$ at site $m$ along $\overrightarrow{n}=(\sin(\theta),0,\cos(\theta)$ reads:
 \be J_{m}(\overrightarrow{n})=\cos^2 (\frac{\theta}{2}) J_{m,+}+\sin^2 (\frac{\theta}{2})J_{m,-}+\sin(\frac{\theta}{2})\cos(\frac{\theta}{2})J^{sf}_{m,+}
 \ee
 where we introduced the spin-flip current at site $m$
 \be  J^{sf}_{m}=\sum_{\sigma}(c^{\dagger}_{m+1,\sigma} c_{m,-\sigma}-c^{\dagger}_{m,\sigma} c_{m+1,-\sigma} ).
 \ee
 Consequently, we are interested in the  the spin current polarized along $\theta=\omega t:$
 \be
 J^{spin}_{m}(\overrightarrow{n})= J_{m}(\overrightarrow{n})- J_{m}(-\overrightarrow{n})
 \ee
 given in terms of Equation (7)  by
  \be
 J^{spin}_{m}(\overrightarrow{n})= (J_{m,+}- J_{m,-})\cos(\theta)+J^{sf}_{m}\sin(\theta).
 \ee

 In Figure 2e) we see the result.  The ring pumps  in the wires an oscillating charge current but a $d.c.$ spin current.
 In this example the magnitude of the field is taken to be such that at normal incidence the ring is pierced by a fluxon
 ($\frac{h}{e}=4.134 *10^{-15}$ MKSA units).  This means that for a 100 atom ring the magnetic field must be of the order
 of $\frac{500}{a^2}$ Tesla, where $a$ is in Angstroms. This is admittedly  too strong a field for the current technology,
 however the field decreases with increasing $N_{ring}$ and, at any rate, the currents are strictly linear with $B$.
 In fact, the arguments of the previous Section concerning the linearity and the secondary role of the spin-orbit interaction remain valid here.\\

  In Figure 3,   a 100 atom ring with spin-orbit interaction ($\alpha_{SO} =0.05$)   rotates  in a constant  external magnetic field. The top left frame  shows the magnetic flux versus  time (in $\frac{\hbar}{t_{h}}$ units).  Below we report the computed  charge current on the first bond of the left lead. The current is measured in electrons per unit time, with the time unit $\frac{\hbar}{t_h}$ appropriate for this model. The current oscillates with the same periodicity as the phase, as one could expect in linear response. In the bottom left position of Fig.3 we see the same current, but refered to the 10th site; there is an evident delay because the disturbance takes time to get there, but then the pattern is the same.

   In  Figure  top right  we show the  spin current on the first bond of the left lead,  keeping the spin quantization axis normal to the rotating ring. It is about $25\%$ of the charge current and follows roughly the same oscillatory pattern. In Figure 3 middle right  we show the spin current on the first bond of the left lead, but now with  the spin quantization axis   parallel to the magnetic field $B$. The spin current  on the first bond of the right lead is exactly opposite to this. There is an evident change of pattern between Figure 2e) and Figure 3d) which is essentially due to the spin-orbit interaction. Finally, Figure 3 bottom right  is like the previous one but with $\alpha_{SO} =0.1$ The intensity of the spin current increases and the oscillations  are wider.

\section{Magnetic field in the plane of the ring: pure spin current }
The next thought experiment  that we propose to produce a spin current requires a time-dependent magnetic field in the plane of the ring. So, $B_{\lfloor}=0$
 and  while there is no magnetic flux, $B_{\shortparallel}=B(t)$ acts on the spin degrees of freedom. The ring is taken in the x-y plane and the spin-polarized current is excited by  a time-dependent  external magnetic field $B(t)$  along the x axis.

Here the assumption that $N_{ring}$ is even is crucial. The analysis of the time evolution  is enormously simplified and can be carried out with generality for any $V(t)$  since the model is bipartite  (i.e. bonds connect sites of two disjoint sublattices), and  can be mapped on a spin-less model which is also bipartite (Fig. 4). A Dirac monopole in the middle (the star in Fig. 4b)) is the point source of a magnetic field which pierces both sub-clusters symmetrically; however it enters the upper ring from below and the lower ring from above. In this way,imparting opposite chirality to the two sub-clusters, it   ensures opposite phase shifts in the corresponding bonds, and this represents the spin-orbit interaction (see Eq. 3). Due to the spin-orbit interaction the parity $P: x \rightarrow -x$ and the reflection $\Sigma$ which sends each sub-cluster to the other fail to commute with $H$, but the product $P\Sigma$ is a symmetry.

 Let us consider sites at the same distance from the ring on both leads  and use the correspondence  $\alpha \rightarrow $ left, spin up; $\beta \rightarrow $ left, spin down;$\gamma \rightarrow $ right, spin up, and $\delta \rightarrow $ right, spin down. Then, the $P\Sigma$ symmetry implies that  at any time, the currents $J$ and the charge densities $\rho$ are constrained by
 \begin{eqnarray}
 J_\alpha (t)=-J_\delta (t)\nonumber\\
 J_\beta (t)= -J_\gamma (t)\nonumber\\
 \rho_\alpha (t)=\rho_\delta(t) \nonumber\\
 \rho_\beta (t)=\rho_\gamma(t) .
  \end{eqnarray}
  In general, the charge on the ring will fluctuate. The question arises: what is the condition for the ring to stay neutral? Introducing the 'and' symbol $\bigwedge$, the continuity equation requires:
  \be J_\alpha (t)=J_\beta (t) \bigwedge J_\gamma (t)=J_\delta (t).\ee Combining with the conditions arising from the $P\Sigma$ symmetry,  one finds
  \be J_\alpha (t)=-J_\gamma (t) \bigwedge  J_\beta (t)=-J_\delta (t),  \ee
  which means, there should be a pure spin current. In general these conditions cannot be met and the ring gets charged, but at half filling these symmetries combined with the charge conjugation symmetry imply that the charge current vanishes identically. This is shown in the Appendix.

\subsection{How the spin current arises}

A vanishing current would trivially satisfy the above symmetries
and theorem in the Appendix. Instead,  the numerical integration
of the Schr\"odinger equation reveals that a spin current flows.
Since this is a novel effect, we must investigate what is the
driving force for the spin current and how one can predict
analytically its magnitude. The perturbation treatment in the
small parameter $\frac{V}{t_{h}}$  although elementary, produces
very large formulas which are of no use here. In order to achieve
a simple  estimate of the effect, and capture the essential
mechanism producing the spin current, we need to introduce  the
concept that the pumping action by a ring on the outside circuit
can be represented by an effective bond. We shall start with the
spinless case and then extend the treatment to the present
problem.
\subsubsection{Effective bond}

  As a preliminary, in order to motivate the renormalised bond idea,
  let us consider the simpler problem of  spinless electrons in the same
  device, but with a normal magnetic field producing a flux in the ring.
  Such a model was studied previously\cite{ciniperfetto}; it was shown
  that by suitable protocols one can insert an integer number of fluxons
  in the ring in such a way that the electronic system in the ring  is
  not left charged and is not excited, while  charge is pumped in the
  external circuit.

   Writing the number current in units of  $\frac{t_{h}}{h}$ it turns out that $\int Jdt$ is of order unity for every fluxon.  In this case, the ring is equivalent to an effective  bond  with hopping $t_{h}\rightarrow t_{h} \exp(i \beta(t)).$ The time dependent vector potential entails the effective  bias across the bond is $e \phi_{eff}=\hbar \dot{\beta}.$   The quantum conductivity of the wire was discussed elsewhere\cite{cini80}; at small $\phi_{eff},$ the number current is $J=-\frac{\phi_{eff}}{\pi \hbar}.$ Integrating over time,
  one finds  that the total charge pumped when a fluxon is swallowed by the ring is $Q=\int{Jdt}={\beta\over\pi}\sim 1$. In other terms, inserting a flux quantum in the ring  we shift an electron in the characteristic hopping time of the system. This simple argument is in good agreement with the numerical results\cite{ciniperfetto}.
\subsubsection{ The driving force}
For the present purposes, we now show that we can  replace the ring by
 a  renormalised bond,
with hopping $t_{h}\rightarrow \tau \exp(i \beta_{\sigma}(t)$ with
$\tau\sim t_{h},$ which implies   an effective  potential drop
across the bond which produces the current. Indeed,  in terms of
the Peierls prescription, this implies  a spin-dependent electric
field $\overrightarrow{E}_{\sigma} $ such that
$\dot{\beta}_{\sigma}=\frac{2\pi}{h}\int
e\overrightarrow{E}_{\sigma} d\overrightarrow{l}$.
   The phase and the  effective potential are spin-dependent and produce
   the spin current.

In  the case with spin, the above approach leads us to change the equivalent model of Figure 4b) to the simplified model of Figure 5, where the vertical bonds again stand for $V(t)$ and the effective bond bears a spin-orbit induced static phase $ \alpha\sim \alpha_{SO}$  which produces no effect at all for  $V$=0. When $V(t)$ is on, however, the electron wave function in the upper wire can interfere with a time dependent contribution from   the opposite spin sector where the phase shift is opposite. Effectively this works like a time dependent phase drop across the upper bond, and an opposite phase drop across the lower one. The spin-dependent electric field is the driving force producing the effect.

 We are now in position to estimate the magnitude of the spin current. In first-order perturbation theory  the amplitude to go from $k_{1+}$ in the upper wire to $k_{2-}$ in the lower wire reads
 \be c_{\alpha}(k_{1}, k_{2},t)=\frac{-i}{\hbar}\int_{0}^{t}d\tau e^{i\omega(k_{2},k_{1})\tau}[1+e^{-2i\alpha+i\D_{12}}] V(\tau)\ee
 where $\omega(k_{2},k_{1})=2(\cos(k_{2})-\cos(k_{1}))$ and $\D_{12}=k_{1}-k_{2}.$
 Since the graph of Fig. 5 is also bipartite, the current $J$  is spin-dependent and site-independent, and  $J\psi_{k}={2 t_{h}\over \hbar}\psi_{k}.$ Therefore the mean current on the top wire is obtained by summing over occupied states:
 \be \langle J_{\alpha} \rangle= {2 t_{h}\over \hbar}\sum_{\k_{1}}^{\cos(k_{1})<0}\sum_{\k_{2}}^{\cos(k_{2})>0}\sin(k_{2})| c_{\alpha}(k_{1}, k_{2},t)|^{2}. \ee
 By definition, the spin current is $J_{s}(\alpha)=J_{\alpha}-J_{-\alpha}.$
 As a simple example, let us take
 \be V(t)=V\theta(t)\theta(T-t).\ee
  Then,
 by taking with $V=\mu_B B$  and then letting $T=\frac{\hbar}{t_{h}}$ (short rectangular  spike)
 we obtain  \be\label{approx} J_{s}(\alpha)\sim \frac{t_{h}}{\hbar}\frac{\sin(\alpha)}{2\pi}(\frac{V T}{\hbar})^2.\ee
 This result is in good agreement with the magnitude of the computed spin currents (see below). We conclude that the effective
 bond concept allows for a simple qualitative picture of the effect.

\begin{figure}[]
\includegraphics*[width=.28\textwidth]{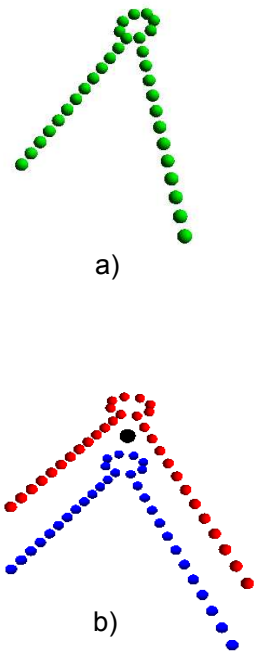}
\caption{Maximally asymmetric connection of the N=6  ring to wires. The circles with  up and down arrows represents the sites; those in the hexagon feel the spin-orbit interaction and the magnetic field. a) Geometry of the device and the magnetic field; all sites are connected horizontally to the first neighbors by spin-diagonal matrix elements, and $B$ flips spins in the hexagon. b) Equivalent cluster for spinless electrons. The star represents the Dirac monopole providing the effective spin-orbit interaction. All sites are connected horizontally to the first neighbors, and those in the ring have also vertical bonds due to the  $V$ magnetic interactions. }
\label{ringgen}
\end{figure}

\begin{figure}[]
\includegraphics*[width=.50\textwidth]{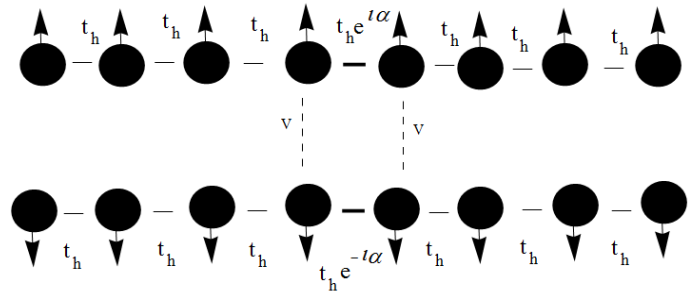}
\caption{The simplified version of our model used to derive Eq. (14). The top (bottom) circles represent the up (down) spin states of a chain; in both horizontal lines the continuous lines stand for identical real hopping matrix elements $t_h$ while the dotted lines represent $t_h e^{i\alpha}$ (top) and $t_h e^{-i\alpha}$ (bottom) connecting, say, sites 0 and 1. The vertical lines stand for $V(t)$ time-dependent hoppings that replace the magnetic interactions in the simplified model. }
\label{ringgen}
\end{figure}

\vspace{2cm}

\begin{figure}[]
\includegraphics*[width=.38\textwidth]{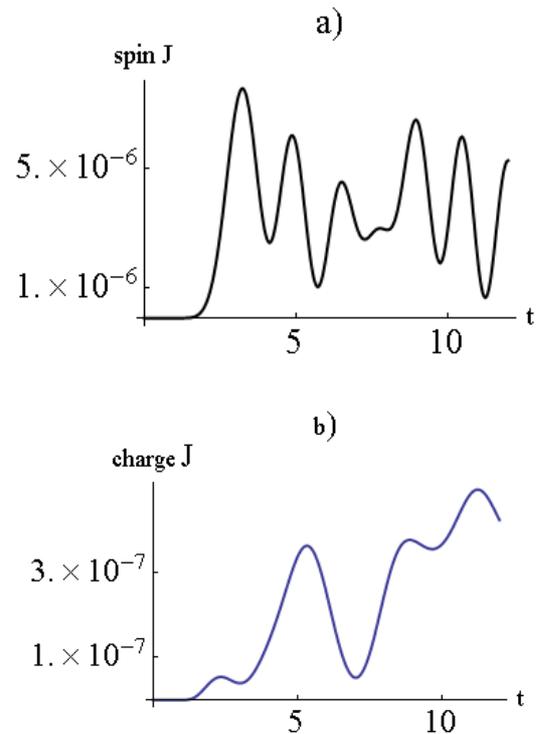}
\caption{Numerical results  of the  model of Eq. (1) with 100 sites in the ring and 100 sites in each lead. The  currents are excited by the sudden switching of a field B=100 Tesla and is quadratic with B. Units are given in the text.
Panel a): spin current $J_{s}$ at half filling and zero temperature. The results hardly change if one takes a temperature of 0.025 $t_{h}$, of the order of room temperature if $t_{h}\sim 1$eV. In both cases the charge current vanishes exactly.
Panel b):   assuming zero temperature and a Fermi energy 0.01$t_{h},$ the spin current does not change, but a charge current  does not vanish any more, although it is still an order of magnitude smaller than the spin current. }
\label{ringgen}
\end{figure}

\vspace{1cm}

\subsection{Field in the plane of the ring: Numerical results}

Test calculations   were  computed for the full model according to
Equation (7). As in the previous Sections,  the  wires are represented in the codes by chains of $N_{wires}$
atoms; if $N_{wires}$ is so large that the results do not change
by increasing $N_{wires}$ we consider that $N_{wires}\rightarrow
\infty $.  The actual $N_{wires}$ which is necessary in a given
calculation increases with the time duration that one wants to
represent. $G^{<}$ was obtained from the determinantal wave
function;  the code evolves the quantum state by a time-slicing
integration of the Schr\"odinger equation.

  The numerical results
strikingly  illustrate the above analytic  findings. In particular, for any time
dependence of $V(t)$ the charge current vanishes identically at
half filling.

\vspace{1cm}

\begin{figure}[]
\includegraphics*[width=.48\textwidth]{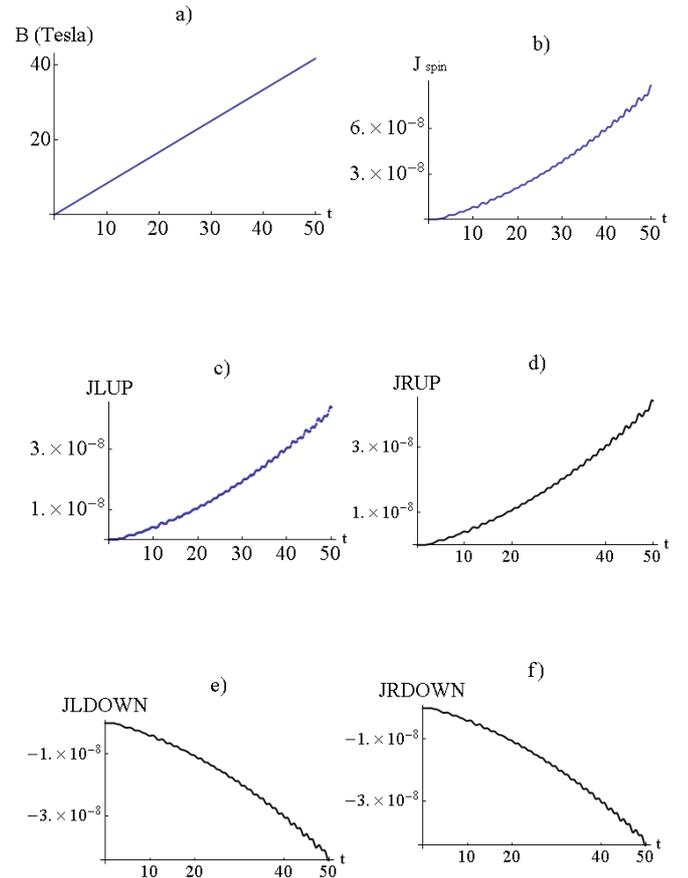}
\caption{ Results of a numerical experiment with an hexagonal ring and  45 sites long wires.a) Time dependence of the in-plane magnetic field; b)spin current on the first site of both wires. To explain  the spin-space distribution of the currents, the computed up spin currents on the first site of left and right wires are reported  in  c) and d) respectively; they are the same. The  spin-down currents in the same sites are reported in (e) and right wire (f)and are the negative of the up-spin currents. This shows that the spin current is even and the charge current vanishes identically.}
\label{ringgen7}
\end{figure}
\vspace{1cm}
In Fig. 6 we present the results\cite{condmat} for the case of a sudden switching of $B$.   We also tested the validity of the simple approximation of  Eq.\ref{approx} for  the full model. For $B=100$  Tesla and $\alpha=1$ one finds $ J_{s}=5*10^{-6}\frac{t_{h}}{\hbar}$ The numerical response to a narrow delta-like  spike yields  $ J_{s}=6*10^{-6}\frac{t_{h}}{\hbar}$ and the quadratic dependence on $B$ is fully confirmed. So the simple approximation works for the full model and although $B=100$  Tesla is high one can easily scale the result to laboratory fields.

 Finite temperatures do not change significantly the results up to $K_{B}T \sim 0.025$ eV. This is interesting since up to now, strongly spin-polarized  currents have been created and detected in ultra-cold atomic gases only\cite{sommer}. Instead, the results are sensitive to the filling. In the lower panel of Figure 6 one can see the result of  raising the Fermi level to $E_{F}=0.01.$ A relatively small charge current develops while numerically the spin current appears to be unaffected by the shift of  $E_{F}.$ As shown at the start of Sect. VI this implies that  the ring gets charged in  the process.

Then, we move to show the numerical results\cite{torino} of our model for  a linear time-dependence of $B$ (Fig 7a)) in a numerical experiment on a hexagonal ring; the infinite wires are  simulated with 45 sites long chains.

Conventionally we take the hopping matrix elements equal to 1 eV and time units are chosen accordingly. In Fig. 7b)
we report the spin current at the first site of each wire (the two are equal since as already remarked spin current is even). The roughly parabolic time dependence is in line with our conclusion that the field dependence is  {\em grosso modo} quadratic in this case (while in the rotating ring experiment above it was linear).  In 7c) and 7e)we report the computed spin-up and spin-down currents in the first site of the left wire, wile in 7d)and 7f) we show the same currents for the right wire. This illustrates the positive parity of both currents and the exact vanishing of the charge current.

In Fig. 8 we use the same model system but with a sinusoidal time dependence of B(t), shown in Fig. 8a). In Fig. 8b) we report the spin-up current response on the 8th site of the left wire; it is identical to the spin-up current on the symmetric site of the right wire and opposite to the spin-down
current on both sides. So, even in this case, a pure spin current
is excited. Interestingly, however, the response of figure 8b) is
not exactly a sine, and a doubled frequency prevails.

\begin{figure}[]
\includegraphics*[width=.44\textwidth]{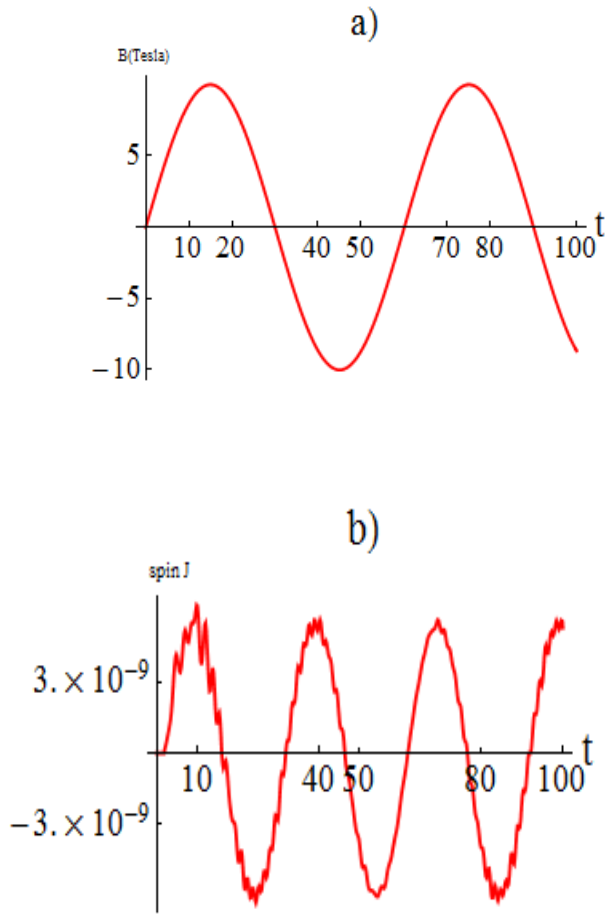}
\caption{a) Sinusoidal  magnetic field, with period $60 \frac{\hbar}{t_{h}}$, used in the second numerical experiment ;b) spin current response on the 8th site of the left wire.}
\label{ringgen8}
\end{figure}

In the above numerical simulations we have illustrated different ways to pump a pure spin current into a
ballistic circuit. The mechanisms advocated in this work do not require magnetic
materials but rely upon either the rotating of the ring in an external magnetic field in the absence of a spin-orbit interaction,
or, as an alternative, upon the combined effects of the spin-orbit
interaction, and a peculiar choice of the geometry that exploits the
maximally asymmetric ring and the external magnetic field.\\

\section{Conclusion and outlook}

We have shown that in principle  by using asymmetrically connected rings at half filling and a time-dependent magnetic field one can pump spin-polarized currents in a ballistic circuit.  Actually there are at least  three novel  different mechanisms available, that after a proper engineering could be of interest for practical applications.

 A rotating   magnetic field piercing  the ring  pumps  an a.c. spin current with partial spin  polarization  and does not rely on the spin-orbit coupling.

 Alternatively one can obtain a d.c. spin current by letting the ring rotate in a static field; the spin-orbit coupling modulates the time dependence but is not required for the spin polarization. The current is proportional to the field.

 The third mechanism achieves the pumping of a purely spin current thanks to both the spin-orbit  interaction, and the action of a tangent time-dependent magnetic field.  The time dependence of the spin current  follows from  the time dependence of the field. The current is approximately  quadratic in the field in this case.

 Ultimately, all these results are consequences of purely quantum effects first discussed in Ref. \cite{ciniperfettostefanucci}, since the classical theory that one finds on Jackson's book\cite{jackson} would not allow any sort of pumping.

  The main message of the present investigation is that by a proper use of quantum rings and magnetic fields one can master the technical problems of spin current handling in future spintronics without the need for nothing else than normal metals. Heavy metals with a large spin-orbit coupling can be needed for some applications involving the last of the three methods proposed here.

\section{Appendix: proof of the pure spin current}

The above findings can be put on a rigorous basis for the sake of the mathematically oriented reader. The study is simplest in the equivalent lattice of Fig. 4b).  Let us consider first any eigenstate  of the instantaneous H, thought of as stationary. Changing sign to all the amplitudes in a sublattice, we get a solution of the one-electron problem with opposite hopping $t_{h}$, and also a solution of the same Schr\"odinger equation with opposite energy. Hence, if $\epsilon$ is an energy eigenvalue, $-\epsilon$ also is, and opposite energy eigenstates must have the same probability on site.   Coming to the many-body state, the sum of the probabilities is exactly one half. In other terms, each site of the equivalent lattice  is exactly half filled, and in the original model the two spin states are exactly half filled on every site. This holds for any static $B$ including  the initial state where  $B=0$. \\

In the adiabatic limit the system is in the instantaneous ground state at each time. Then, no charge current is allowed, because  the total occupation of each site in Fig. 4a) is fixed; moreover no spin current is allowed either, since it would alter  the  occupation of the sites in Fig. 4b), which is also bipartite. Therefore the adiabatic evolution of this system is trivial. Since we are interested in the non-adiabatic evolution, the beautiful analysis by Avron et al.\cite{adiabatic} does not apply here.\\

Next, we consider the tome evolution in the presence of the time-dependent field.
To show that during the time evolution  $B(t)$ produces a pure spin current in the half filled  system, we change to a staggered spin-reversed  hole representation with $c^{\dagger}_{i,\sigma}=s b_{i,-\sigma},$ where $s=1$ in one sublattice and $s=-1$ in the other; here the operators refer to any site of the device. The transformation of any bond goes as follows:
$t^{\sigma}_{n,m}c^{\dagger}_{n,\sigma}c_{m,\sigma} \rightarrow -t^{\sigma}_{n,m}b_{n,-\sigma}b^{\dagger}_{m,-\sigma}$. Since $t^{\sigma}$ is Hermitean this is the same as   $t^{\sigma *}_{m,n}b^{\dagger}_{m,-\sigma}b_{n,-\sigma}$ and since opposite spins have conjugate hoppings  we may rewrite this as $t^{-\sigma}_{m,n}b^{\dagger}_{n,-\sigma}b_{n,-\sigma}.  $ On the other hand, in the ring the sites coupled by $V$ belong to opposite sublattices and so the transformation gives:

\be H_{B}\rightarrow V(t) \sum_{i \in ring}(b^{\dagger}_{i,\uparrow}b_{i,\downarrow}+ b^{\dagger}_{i,\downarrow}b_{i,\uparrow}   ).\ee

In this way,  at every time $t$ the transformed hole Hamiltonian $\tilde{H}(b,b^{\dagger})$ depends on $b$ operators exactly as the original Hamiltonian
 $H(c,c^{\dagger})$ depends on the  $c$ operators. In both pictures the evolution starts in the ground state at half filling  and evolves in the same way. Therefore,  at any time and   for any site $n$,
 \be \langle b^{\dagger}_{n,\sigma}(t) b_{n,\sigma}(t)\rangle_{b} = \langle c^{\dagger}_{n,\sigma} (t) c_{n,\sigma}(t)\rangle_{c}. \ee Here  the l.h.s. is the average at time $t$ in the $b$ picture while the r.h.s. is averaged  at time $t$ in the $c$ picture.
 Hence, operating the canonical transformation on the l.h.s.,
\be  \langle 1-n_{n,-\sigma}(t)\rangle_{c} = \langle n_{n,\sigma}(t)\rangle_{c}, \ee
which implies that the mean total occupation of each site is conserved. This cannot be true if charge currents exist. Indeed, let us consider  the operators straddling  each bond: since at each time
 \be \langle b^{\dagger}_{n+1,\sigma} b_{n,\sigma}\rangle_{b} = \langle c^{\dagger}_{n+1,\sigma} c_{n,\sigma}\rangle_{b} \ee
we may conclude that
 \be \langle c^{\dagger}_{n,-\sigma} c_{n+1,-\sigma}\rangle_{c} = \langle c^{\dagger}_{n+1,\sigma} c_{n,\sigma}\rangle_{c}. \ee
Hence the current is pure spin current, q.e.d.

\section{Acknowledgements} The authors are grateful to Matteo Colonna for help with a computer code during the early stages of this project.
%


\end{document}